\documentclass[%
 reprint,
 superscriptaddress,
 amsmath,amssymb,
 aps,
]{revtex4-1}

\usepackage{extarrows}
\usepackage{graphicx}
\usepackage{dcolumn}
\usepackage{bm}
\usepackage{xcolor}
\usepackage{url}
\usepackage{longtable}

\usepackage{subfigure}
\usepackage{multirow}

\begin{document}


\title{Vector meson-nucleon scattering length $|\alpha_{VN}|$ and trace anomalous energy contribution to the nucleon mass $T_{A}$}

\email{These authors contributed equally: Chengdong Han and Wei Kou.}

\author{Chengdong Han}
\email{chdhan@impcas.ac.cn}
\affiliation{Institute of Modern Physics, Chinese Academy of Sciences, Lanzhou 730000, China}
\affiliation{University of Chinese Academy of Sciences, Beijing 100049, China}

\author{Wei Kou}
\email{kouwei@impcas.ac.cn}
\affiliation{Institute of Modern Physics, Chinese Academy of Sciences, Lanzhou 730000, China}
\affiliation{University of Chinese Academy of Sciences, Beijing 100049, China}

\author{Rong Wang}
\email{rwang@impcas.ac.cn}
\affiliation{Institute of Modern Physics, Chinese Academy of Sciences, Lanzhou 730000, China}
\affiliation{University of Chinese Academy of Sciences, Beijing 100049, China}

\author{Xurong Chen}
\email{xchen@impcas.ac.cn (Corresponding author)}
\affiliation{Institute of Modern Physics, Chinese Academy of Sciences, Lanzhou 730000, China}
\affiliation{University of Chinese Academy of Sciences, Beijing 100049, China}
\affiliation{Guangdong Provincial Key Laboratory of Nuclear Science, Institute of Quantum Matter, South China Normal University, Guangzhou 510006, China}



\begin{abstract}
  Low-energy scattering processes of vector meson and nucleon are an important window for studying non-perturbative QCD.
  The interaction of vector meson with nucleon, vector meson-nucleon scattering length $|\alpha_{VN}|$, is an important component of the study of hadronic interactions. 
  Nowadays many scattering length values $|\alpha_{VN}|$ have been reported using the recent photoproduction experiment data or quasi data.
  In addition, the study of trace anomalous energy contribution to the proton mass is also a hot topic in non-perturbative QCD and hadron physics.
  However, it is difficult to measure proton trace anomalous energy experimentally, and the study of the trace anomaly of proton is still inconclusive.
  In this study, we established the relationship between the scattering length of the vector meson-proton $|\alpha_{Vp}|$ and the trace anomaly contribution of the proton mass $T_{A}$.
  With the scattering length values extracted by using the Vector Meson Dominance model, we obtained the trace anomaly contribution of the proton mass $T_{A}$ = (22.8$\%$ $\pm$ 1.2$\%$),
  which is of similar order of magnitude as the 23$\%$ given by Lattice QCD calculation.
  We conjecture that the trace anomaly contribution of nucleon is independent of the type of vector meson probe.
  We hope that high precision measurements of vector meson-nucleon scattering length could give us a better chance to explore the origin of the nucleon mass.
\end{abstract}

\maketitle


\section{Introduction}
\label{introduction}
Hadronic interactions, hadron internal structure and dynamical hadron-mass generation are the hot research fields in the non-perturbative quantum chromodynamics (QCD).
Since the discovery of the vector mesons, vector mesons have become useful probe for the study of hadronic matter and hadronic interaction.
Experimentally, the vector meson-nucleon interaction can be investigated using vector meson photoproduction within the Vector Meson Dominance (VMD) model \cite{SAKURAI19601}.
Since the vector meson-nucleon interaction is an important area of investigation in the non-perturbation domain of QCD,
the absolute value of the scattering length can be determined from the total near-threshold vector-meson photoproduction cross-section \cite{Gell-Mann:1961jim}
or the differential cross-section of vector-meson photoproduction at threshold \cite{Titov:2007xb} based on the theoretical model.
At present, the $\omega p$, $\omega n$, $\rho^{0}p$, $\phi p$, $J/\psi p$, $\psi(2S)p$ and $\Upsilon$p scattering lengths have been fully analysed in Refs. \cite{Strakovsky:2014wja,Strakovsky:2019bev,Strakovsky:2020uqs,Pentchev:2020kao,Wang:2022xpw,Wang:2022zwz,CDHan:2022cdhan,Strakovsky:2021vyk}. 
In addition, the measurement of proton trace anomalous energy is always a challenge in the experiments.
Theoretically, the QCD interpretation of proton trace anomaly is still ambiguous. To connect the theory with the experiment for the nucleon trace anomalous energy, the authors
 \cite{Wang:2019mza,Kou:2021bez} recently extracted the trace anomaly by analyzing the near-threshold photoproduction data of $\phi$ and $J/\psi$ vector mesons, which give ones a method to estimate the proton mass decomposition \cite{Ji:1995sv}. The proton mass is an endogenous property and is generally considered not to vary with the type of probe. The various components of the decomposed proton mass should also follow this principle and should be consistent within a certain error range. To study the nature of the trace anomaly energy of the proton, we consider the near-threshold photoproduction process of the vector mesons. We start from the vector meson-nucleon scattering length and expect a uniform nucleon trace anomaly contribution based on different meson probe experiments.
 
In this work, we extend our previous works \cite{Wang:2019mza,Kou:2021bez,CDHan:2022cdhan} and relate two potential observables of the near-threshold photoproduction process, the scattering length and the proton trace anomaly energy. We find a new method for determining the nucleon trace anomaly energy, which requires the vector meson-nucleon scattering length as inputs. The paper is organized as follows. In Sec. \ref{sec:scattering length and diffxsection}, we briefly describe the scattering length of vector meson-nucleon which obtained from the VMD model. We then discuss the method of extracting proton trace anomalous contributions under the VMD model and relate it to the scattering length. Our main results are discussed in Sec. \ref{sec:scattering length and TA}. Finally, we display some discussion and summary.

\section{Vector meson-nucleon scattering length $|\alpha_{VN}|$ from differential cross sections}
\label{sec:scattering length and diffxsection}
Within the VMD model, the total $\gamma N\to VN$ cross-section is related to both the total $VN \to VN$ cross-section at threshold energy and
the scattering length $\alpha_{VN}$ by \cite{Titov:2007xb}:
\begin{equation}
  \begin{split}
    & \sigma^{\gamma N}\left(s_{t h r}\right) = \frac{\alpha \pi}{\gamma_{V}^{2}} \frac{q_{VN}}{k_{\gamma N}} \cdot \sigma^{VN}\left(s_{t h r}\right) =\frac{\alpha \pi}{\gamma_{V}^{2}} \frac{q_{VN}}{k_{\gamma N}} \cdot 4 \pi \alpha_{VN}^{2} \\
    \label{eq:total}
  \end{split}
\end{equation}
where $\alpha = 1/137$ is the fine structure constant, and $V$ is a index which represents the vector mesons (e.g. $\omega$, $\rho^{0}$, J/$\psi$, etc.)
The $k_{\gamma N}$ and $q_{VN}$ in the above equation are the momenta in the center-of-mass of the initial and final state particles, respectively,
and $\gamma_{V}$ is the photon-vector meson coupling constant obtained from the $V \to e^+e^-$ decay width. 
Eq.\ref{eq:total} is taken at the threshold energy, where s$_{thr}$ = $(M + m)^{2}$ with M and m being the masses of the vector meson and nucleon, respectively.

In order to estimate the scattering length with the experimental data of the differential photoproduction cross section $d\sigma^{\gamma N}/dt$,
Pentchev and Strakovsky \cite{Pentchev:2020kao} established the relation between the total and differential cross sections at threshold.
The total cross section is defined as an integral over the interval t [t$_{min}$(s), t$_{max}$(s)],
\begin{equation}
    \begin{split}
      & \sigma^{\gamma N}(s) = \int_{t_{min}}^{t_{max}} \frac{d \sigma^{\gamma N}}{dt}(s,t)dt \\
      & \xlongequal{t_{min} \rightarrow t_{max}} \Delta t \frac{d \sigma^{\gamma N}}{dt}(s_{thr},t_{thr}) \\
      & = 4q_{VN}k_{\gamma N} \frac{d \sigma^{\gamma N}}{dt}(s_{thr}, t_{thr}).
      \label{eq:diffcrosec_with_totalcrosssec}
    \end{split}
\end{equation}
When approaching threshold $t_{min}$ $\rightarrow$ $t_{max}$ in Eq. (\ref{eq:diffcrosec_with_totalcrosssec}),
the relationship between the total and differential cross sections at threshold is established.
Where the $\Delta t$  = $\left|t_{max} - t_{min} \right|$ = 4$q_{VN}k_{\gamma N}$ and $t_{thr} = t_{min}(s_{thr}) =  t_{max}(s_{thr}) = -M^{2}m/(M+m)$.
Combining Eq. (\ref{eq:diffcrosec_with_totalcrosssec}) and Eq. (\ref{eq:total}), the relationship between the scattering length and the differential cross-section is expressed as:
\begin{equation}
    \begin{split}
        \frac{d \sigma^{\gamma V}}{d t}\left(s_{t h r}, t=t_{thr}\right)=\frac{\alpha \pi}{\gamma_{V}^{2}} \frac{\pi}{k_{\gamma N}^{2}} \cdot \alpha_{ VN}^{2}, \\
	\label{eq:diffxsection}
    \end{split}
\end{equation}
A key problem in determining the scattering length at threshold t$_{thr}$ is to extrapolate the cross section to the point of
t$\rightarrow$t$_{thr}$ or s$\rightarrow$s$_{thr}$.

\begin{equation}
    \begin{split}
        \frac{d \sigma^{\gamma V}}{d t}\left(s_{t h r}, t=0\right)=\frac{\alpha \pi}{\gamma_{V}^{2}} \frac{\pi}{k_{\gamma N}^{2}} \cdot \alpha_{ VN}^{2}.   
	\label{eq:diffxsection_t0}
    \end{split}
\end{equation}

The $d\sigma^{\gamma V}/dt(s_{thr}, t=0)$ at left-hand side of Eq. (\ref{eq:diffxsection_t0}) is not a directly measurable quantity,
as it requires extrapolation of the energy to the threshold and extrapolation of t from the physical region ($t_{min}<t<t_{max}$) to the non-physical point t = 0.
Therefore, when the vector meson nucleon scattering length is extracted from the differential cross-section data, we can not only extrapolate the energy at the threshold, but also extrapolate t to t = 0.

The following exponential function was used to fit the differential cross-section measurements of near-threshold vector meson photoproductions
on hydrogen target or deuterium target.
\begin{equation}
\begin{split}
\frac{d\sigma}{dt}=Ae^{-bt},
\end{split}
\label{eq:exp_fit}
\end{equation}
where $A=d\sigma/dt|_{t=0}$ denotes the forward differential cross-section and $b$ describes the slope parameter.
Combining Eqs. (\ref{eq:diffxsection}), (\ref{eq:diffxsection_t0})
and (\ref{eq:exp_fit}), the forward differential cross-section $d\sigma/dt|_{t=0}$ or $d\sigma/dt|_{t=thr}$
can be obtained, and furhter extract the vector meson-nucleon scattering length $\alpha_{VN}$.

A dispersive analysis to extract the $\Upsilon-p$ scattering length from $\gamma p$ $\rightarrow$ $\Upsilon p$ experiments was presented
by Gryniuk et al. \cite{Gryniuk:2020mlh}.
For their framework, the imaginary part of the $\Upsilon-p$ forward scattering amplitude $T_{\Upsilon p}$ determined by $\gamma p$ $\rightarrow$ $\Upsilon p$
cross section measurements, and the real part of the scattering amplitude $T_{\Upsilon p}$ is obtained from a once-subtracted dispersion relation.
The subtraction constant related to $\Upsilon-p$ scattering amplitude $T_{\Upsilon p}$ is determined by fitting the $\gamma p$ $\rightarrow$ $\Upsilon p$
differential photoproduction cross section data at t = 0, as follows
\begin{equation}
  \begin{aligned}
     & \frac{d\sigma_{\gamma p \rightarrow \Upsilon p}}{dt}\bigg|_{t=0} 
      = (\frac{ef_{\Upsilon}}{M_{\Upsilon}})^{2} \frac{1}{64 \pi s q^{2}_{\gamma p}} |T_{\Upsilon p}|^{2},
  \end{aligned}
  \label{eq:forward_cross_section_TVN}
\end{equation}
where $f_{\Upsilon}$ is the $\Upsilon$ decay constant, and $q_{\gamma p}$ represents the magnitude of the
photon three-momentum in the center mass frame of the photoproduction process of $\Upsilon$ meson.
The real part of the forward scattering amplitude at threshold $T_{VN}$ in Eq.(\ref{eq:forward_cross_section_TVN})
is directly related to the $V-N$ scattering length $\alpha_{VN}$ as \cite{Gryniuk:2020mlh}
\begin{equation}
\begin{split}
T_{VN}=8\pi(M+m)\alpha_{VN}.
\end{split}
\label{eq:TVN}
\end{equation}

\section{Scattering lengths $|\alpha_{VN}|$ and the trace anomaly contribution of nucleon mass $T_{A}$}
\label{sec:scattering length and TA}
Understanding how the hadron mass emerges in QCD is of utmost importance. An effective approach is to consider the contribution of the quark and gluon that are the basic unit of nucleon.
How quarks and gluons dynamically produce the whole nucleon mass is a very fundamental question. In Refs. \cite{Ji:1994av,Ji:1995sv}, Ji first defined the proton mass decomposition with QCD Hamiltonian operators and assumed that the hadron mass is
calculated as the expectation value of the Hamiltonian at the
hadron rest frame:
\begin{equation}
	M_N=\left.\frac{\left\langle P\left|H_{\mathrm{QCD}}\right| P\right\rangle}{\langle P \mid P\rangle}\right|_{\text {rest frame }},
	\label{eq:Hardonstate}
\end{equation}
which is decomposed into four terms characterized by
the QCD trace anomaly parameter $b(\mu^2)$
and the momentum fraction $a(\mu^2)$ carried by all quarks \cite{Ji:1995sv}.
The four terms of the proton mass partitions are written as \cite{Ji:1995sv},
\begin{equation}
	\begin{aligned}
		&M_{q} =\frac{3}{4}\left(a-\frac{b}{1+\gamma_{m}}\right) M_{N}, \ \
		M_{g} =\frac{3}{4}(1-a) M_{N}, \\
		&M_{m} =\frac{4+\gamma_{m}}{4\left(1+\gamma_{m}\right)} b M_{N}, \ \
		M_{a} =\frac{1}{4}(1-b) M_{N},
	\end{aligned}
	\label{eq:Protom_mass}
\end{equation}
where the anomalous dimension of quark mass $\gamma_{m}$ \cite{Buras:1979yt} describes the renormalization information, and $b$ denotes the gluon trace anomaly parameter.
The $a(\mu^2)$ of the quarks is calculated 
with all the quark distributions determined by experimental measurements, as follows
\begin{equation}
	a\left(\mu^{2}\right)=\sum_{f} \int_{0}^{1} x\left[q_{f}\left(x, \mu^{2}\right)+\bar{q}_{f}\left(x, \mu^{2}\right)\right] d x.
	\label{eq:Momentum_frac}
\end{equation}
The first three terms in Eq. (\ref{eq:Protom_mass}) can be easily 
understood with the classical field theory.
However the last term is an extension of the classical description 
in the quantum field theory -- the quantum anomaly \cite{Kou:2021bez}.

With the VMD model, the forward differential cross section of the vector meson $V$ ($\omega$, $\rho^{0}$, $\phi$, J/$\psi$, etc.) photoproductions
on the proton target is followed as,
\begin{equation}
  \begin{aligned}
     & \frac{d\sigma_{\gamma N \rightarrow VN}}{dt}\bigg|_{t=0} \\
     & = \frac{3\Gamma(V \rightarrow e^{+}e^{-})}{\alpha m_V} (\frac{k_{VN}}{k_{\gamma N}})^{2} \frac{d\sigma_{VN \rightarrow VN}}{dt} \bigg|_{t=0} \\
     & = \frac{3\Gamma(V \rightarrow e^{+}e^{-})}{\alpha m_V} (\frac{k_{VN}}{k_{\gamma N}})^{2} \frac{1}{64 \pi} \frac{1}{m^{2}_{V}(\lambda^{2}-m^{2}_{N})}|F_{VN}|^{2}
  \end{aligned}
  \label{eq:forward_cross_section}
\end{equation}
where $\alpha$ = 1/137 is the fine structure constant, $k^{2}_{ab}$ denotes the center-of-mass momentum square of the corresponding two-body system, 
$\Gamma$ is the partial decay width of the $V \rightarrow e^{+}e^{-}$,
$\lambda=(p_{N}p_{V}/m_{V})$ is the nucleon energy at the quarkonium rest frame \cite{Kharzeev:1998bz}, and $F_{VN}$ represents the invariant
amplitude of $V-N$ elastic scattering \cite{Kharzeev:1995ij,Kharzeev:1998bz}.

In order to determine the trace anomaly contribution of proton mass, 
the previous analysis method is to obtain the trace anomaly contribution of proton mass by fitting the experimental data of vector meson photoproductions near the threshold \cite{Kou:2021bez,Wang:2019mza}.
With the Refs. \cite{Kharzeev:1995ij,Kharzeev:1998bz}, the invariant amplitude of $V-N$ elastic scattering takes the form 
\begin{equation}
  \begin{aligned}
    F_{V N} & \simeq r_{0}^{3} d_{2}  \frac{8 \pi^{2}M_Nm_V}{27}\left(M_{N}-
    \left\langle N\left|\sum_{i=u, d, s} m_{i} \bar{q}_{i} q_{i}\right| N\right\rangle\right) \\
    & = r_0^3d_2\frac{8\pi^2}{27}(1-b)M_N^2m_V.
  \end{aligned}
  \label{eq:amplitude}
\end{equation}
Taking the forward limit ($t\to0$) and comparing with Eq. (\ref{eq:TVN}) we have
\begin{equation}
  \begin{split}
    |\alpha_{VN}|=\frac{4\pi d_{n}^{1S}m^{2}_{N}m_{V}r^{3}_{0}T_{A}}{27\sqrt{s_{thr}}},
  \end{split}
  \label{eq:avp_vs_mass_TA_equ}
\end{equation}
where $|\alpha_{VN}|$ denotes the vector meson-nucleon scattering lengths, 
$m_{N}$ is the nucleon mass, and $m_{V}$ is the vector meson mass.
$T_{A}$ is the trace anomalous energy contribution to the nucleon mass, which is expressed as $T_A=(1-b)/4$.
The “Bohr” radius $r_{0}$ of the vector meson $V$ is given by \cite{Kharzeev:1995ij},
\begin{equation}
\begin{split}
  r_{0}=\frac{4}{3\alpha_{s}}\frac{1}{m_{q}}
\end{split}
\label{eq:equ_r0}
\end{equation}
where $m_{q}$ represents the constituent mass of quark and $\alpha_{s}$ is the running coupling.
In this study, we choose the “Bohr” radius size with $r_{0}(\omega)$  = 0.75 fm,
$r_{0}(\rho)$ = 0.75 fm \cite{Krutov:2016uhy,Bhagwat:2006pu,Grigoryan:2007my}, $r_{0}(\phi)$ = 0.41 fm, $r_{0}(J/\psi)$ = 0.20 fm \cite{Kharzeev:1995ij}, and $r_{0}(\Upsilon)$ = 0.10 fm. 
Where $r_{0}(J/\psi)$, $r_{0}(\phi)$ and $r_{0}(\Upsilon)$ radii are determined using the ``Riedberg" energy of the quark-antiquark pairs. 
For the charmonium ground state J/$\psi$, a naive estimate of the ``Rydberg" energy $E_{J/\psi}$ is $m_D+m_{\bar{D}}-m_{J / \psi}$, which means that the $c\bar{c}$ pair will be pulled apart to generate the $D\bar{D}$ pair.
The above relation could be used to obtain the J/$\psi$'s ``Bohr" radius by $E_{J / \psi}=\left(1 / m_c r_{J / \psi}^2\right)$ (see Refs. \cite{Kharzeev:1995ij,Wang:2019mza} for details). Although the selection of the ``Bohr" radius is not very rigorous, we explained the validity of the above radius values and analyzed the uncertainties associated with the selection of the ``Bohr" radius in Ref. \cite{Kou:2021bez}. 
The Wilson coefficient $d_{n}^{1S}$ is found in Refs. \cite{Kharzeev:1995ij,Peskin:1979va,Kharzeev:1996tw} as
\begin{equation}
\begin{split}
  d_{n}^{1S}=(\frac{32}{N_{c}})^{2}\sqrt{\pi}\frac{\Gamma(n+\frac{5}{2})}{\Gamma(n+5)},
\end{split}
\label{eq:equ_dn}
\end{equation}
where N$_{c}$ = 3 is the number of colors. Eq. (\ref{eq:avp_vs_mass_TA_equ}) combines the information of vector meson-nucleon scattering length with the trace anomalous energy contribution to the nucleon mass, the later cannot be directly determined by experimental measurements.

\begin{figure}[htp]
\centering
\includegraphics[width=0.47\textwidth]{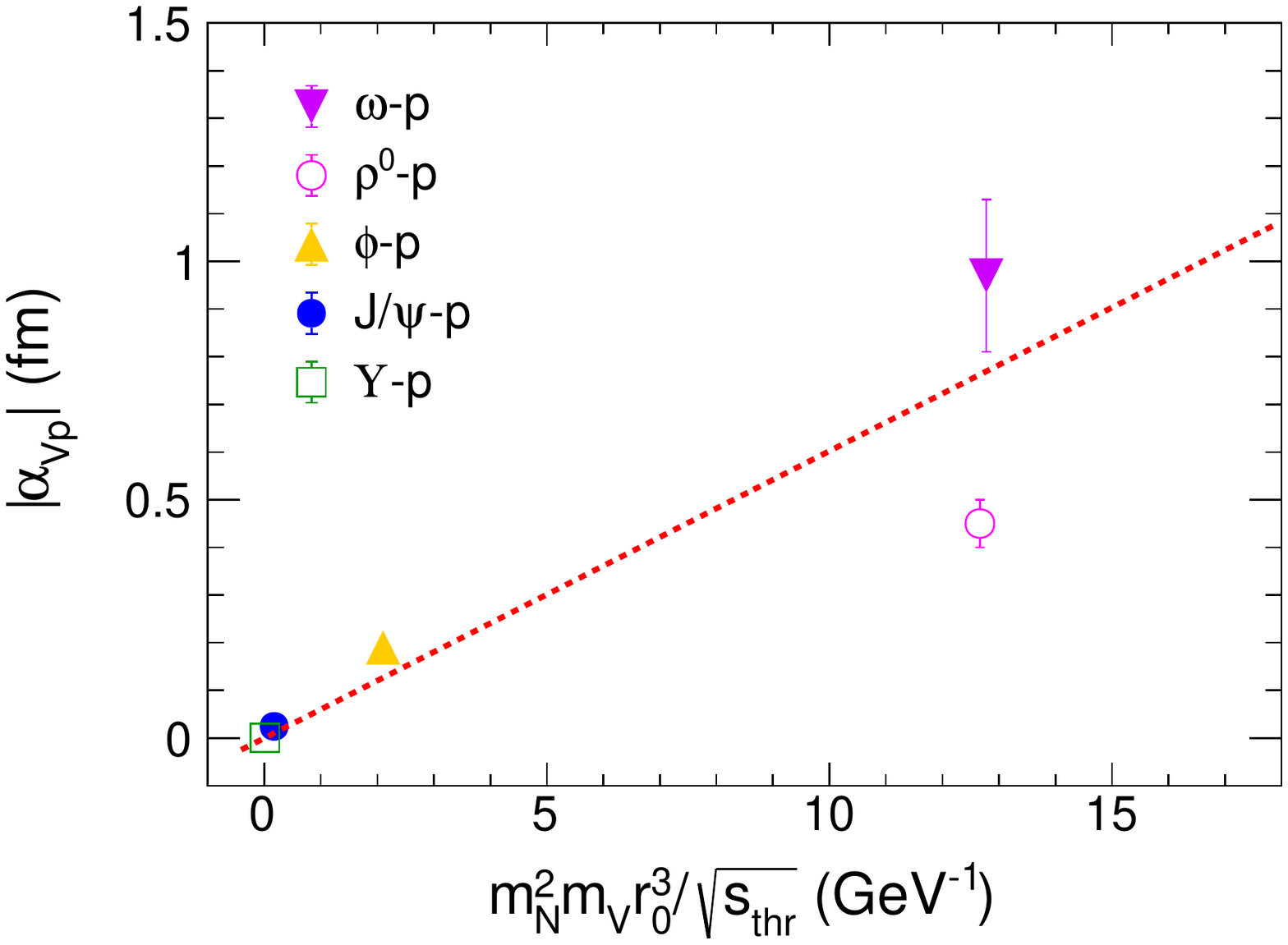}
\caption{
  The relationship between the scattering lengths $|\alpha_{VN}|$ as a function of the $m_{N}^{2}m_{V}r_{0}^{3}/\sqrt{s_{thr}}$ of the vector mesons,
  including $\omega$, $\rho^{0}$, $\phi$, J/$\psi$ and $\Upsilon$ mesons.
}
\label{fig:avp_vs_mass_TA}
\end{figure}

Table \ref{tab:avp_scatteringlength_vari} shows the value of scattering length $|\alpha_{Vp}|$ with $\omega$p \cite{Ishikawa:2019rvz}, $\rho^{0}$p \cite{Wang:2022zwz,klein1997appear}, $\phi$p \cite{Strakovsky:2020uqs,CLAS:2013jlg,Dey:2014tfa}, $J/\psi$p \cite{Pentchev:2020kao} and $\Upsilon$p \cite{Strakovsky:2021vyk} mesons
and the value of ``Bohr'' radius, and threshold energy of different vector meson-proton interaction.
According to Wang's \cite{Wang:2022zwz} previous work, we extracted the $\rho^{0}$p scattering length from the differential cross-section data \cite{klein1997appear} of near-threshold $\rho^{0}$ photoproductions in t to t = 0, which is used for this work.
In addition, the differential cross-section $\phi$-meson photoproduction data \cite{CLAS:2013jlg,Dey:2014tfa} from the CLAS threshold measurements are used for evaluating the $\phi$p scattering length $|\alpha_{\phi p}|$ in t to t = 0,
and the extracted $\alpha_{\phi p}$ is used for this analysis.
The Fig. \ref{fig:avp_vs_mass_TA} shows the relationship between the vector meson-proton scattering length and the trace anomaly contribution of the proton mass. 
By fitting the distribution in Fig. \ref{fig:avp_vs_mass_TA} with a linear function, the fitting result is $\frac{4\pi d_{2}^{1s}}{27}T_{A}$  = (0.060 $\pm$ 0.003).
Since $\frac{4\pi d_{2}^{1S}}{27}$ is a constant quantity equal to 1.333, $T_{A}$ = (0.228 $\pm$ 0.012) is calculated.

\begin{table*}
  \caption{
    The values of scattering length $|\alpha_{Vp}|$ of $\omega$p, $\rho^{0}$p, $\phi$0, J/$\psi$p and $\Upsilon$p
    ,the value of ``Bohr'' radius $r_{0}$, and threshold energy $\sqrt{s_{thr}}$ of different vector meson-nucleon interaction.
  }
  \begin{center}
    \begin{ruledtabular}
      \begin{tabular}{ cccccccccc }
         Vector meson  & m$_{V}$ (GeV) & $|\alpha_{Vp}|$ (fm)                                         &  r$_{0}$ (fm)     & $\sqrt{s_{thr}}$ (GeV)  \\
        \hline
        $\omega$       &  0.7827    & 0.97 $\pm$ 0.16 \cite{Ishikawa:2019rvz}                         &  0.75           &  1.721  &\\
        $\rho^{0}$     &  0.7753    & 0.45 $\pm$ 0.05 \cite{Wang:2022zwz,klein1997appear}                             &  0.75           &  1.713  &\\
        $\phi$         &  1.0195    & 0.19 $\pm$ 0.01 \cite{Strakovsky:2020uqs,CLAS:2013jlg,Dey:2014tfa}                          &  0.41            &  1.958  &\\
        J/$\psi$       &  3.0960    & 0.0245 $\pm$ 0.039 \cite{Pentchev:2020kao}                      &  0.20           &  4.035  &\\
        $\Upsilon$     &  9.4603    & (0.51 $\pm$ 0.03) $\times$ 10$^{-3}$ \cite{Strakovsky:2021vyk}  &  0.10           &  10.40  &\\
      \end{tabular}
    \end{ruledtabular}
  \end{center}
  \label{tab:avp_scatteringlength_vari}
\end{table*}

\section{Discussion and summary}
\label{Discussion and summary}
In the study, we established the relationship between the vector meson-nucleon scattering length and the trace anomaly contribution of the nucleon mass
for the first time through the differential cross section of vector meson photoproductions process at near threshold.
This means that as long as the scattering length of the vector meson-proton is accurately measured of near-threshold meson photoproductions,
the proton trace anomaly energy can be obtained with the relationship we obtained.
The contribution of trace anomaly in proton we extracted is (22.8$\%$ $\pm$ 1.2$\%$), which is consistent with the result of lattice QCD calculation \cite{Yang:2018nqn}.
The percentage of trace anomaly energy inside the proton $T_{A}$ we obtained depends on the “Bohr” radius $r_{0}$ of vector meson from the theoretical model calculations,
and the extraction value of scattering lengths $|\alpha_{VN}|$ from the differential cross-section data of near-threshold meson photoproductions.
A related analysis of systematical uncertainties can be found in Ref. \cite{Kou:2021bez}.
As we knows, the origin of the mass of proton is very complicated in modern particle physics and no definite conclusion has been made about it in the recent years.
In addition, it is very difficult to measure the proton trace anomalous energy in experiments.
While different vector meson probes lead to differences in the specific process of near-threshold photoproduction as well as scattering length,
we can still glimpse that the components of the proton mass are independent of the specific reaction process.
In this work, we aim to understand the near-threshold photoproduction of vector mesons from other perspectives
and to find a experimental measurement quantity as the input for the study of the origin of the proton mass.
The above result of trace anomaly contribution in proton may provide useful theoretical information for an in-depth understanding
of nucleon interaction with vector mesons and the trace anomalous energy contribution to the nucleon mass.

Furthermore, the Electron-Ion Collider in the USA (EIC) \cite{Accardi:2012qut}
and the Electron-ion collider in China (EicC) \cite{Chen:2018wyz,Chen:2020ijn,Anderle:2021wcy} will provide a favourable circumstances
to study the near-threshold vector meson photoproduction by exploiting the virtual photon flux.
The vector meson photoproduction experiments at EIC and EicC will be further test the VMD model
and will also strengthen our understanding on the properties of hadronic matter and hadronic interactions.

\begin{acknowledgments}
This work is supported by the Strategic Priority Research Program of Chinese Academy of Sciences
under the Grant NO. XDB34030301, the National Natural Sci- ence Foundation of China No. 12005266
and Guangdong Major Project of Basic and Applied Basic Research No. 2020B0301030008.
\end{acknowledgments}

\bibliographystyle{apsrev4-1}
\bibliography{refs.bib}

\begin{thebibliography}{33}%
\makeatletter
\providecommand \@ifxundefined [1]{%
 \@ifx{#1\undefined}
}%
\providecommand \@ifnum [1]{%
 \ifnum #1\expandafter \@firstoftwo
 \else \expandafter \@secondoftwo
 \fi
}%
\providecommand \@ifx [1]{%
 \ifx #1\expandafter \@firstoftwo
 \else \expandafter \@secondoftwo
 \fi
}%
\providecommand \natexlab [1]{#1}%
\providecommand \enquote  [1]{``#1''}%
\providecommand \bibnamefont  [1]{#1}%
\providecommand \bibfnamefont [1]{#1}%
\providecommand \citenamefont [1]{#1}%
\providecommand \href@noop [0]{\@secondoftwo}%
\providecommand \href [0]{\begingroup \@sanitize@url \@href}%
\providecommand \@href[1]{\@@startlink{#1}\@@href}%
\providecommand \@@href[1]{\endgroup#1\@@endlink}%
\providecommand \@sanitize@url [0]{\catcode `\\12\catcode `\$12\catcode
  `\&12\catcode `\#12\catcode `\^12\catcode `\_12\catcode `\%12\relax}%
\providecommand \@@startlink[1]{}%
\providecommand \@@endlink[0]{}%
\providecommand \url  [0]{\begingroup\@sanitize@url \@url }%
\providecommand \@url [1]{\endgroup\@href {#1}{\urlprefix }}%
\providecommand \urlprefix  [0]{URL }%
\providecommand \Eprint [0]{\href }%
\providecommand \doibase [0]{http://dx.doi.org/}%
\providecommand \selectlanguage [0]{\@gobble}%
\providecommand \bibinfo  [0]{\@secondoftwo}%
\providecommand \bibfield  [0]{\@secondoftwo}%
\providecommand \translation [1]{[#1]}%
\providecommand \BibitemOpen [0]{}%
\providecommand \bibitemStop [0]{}%
\providecommand \bibitemNoStop [0]{.\EOS\space}%
\providecommand \EOS [0]{\spacefactor3000\relax}%
\providecommand \BibitemShut  [1]{\csname bibitem#1\endcsname}%
\let\auto@bib@innerbib\@empty
\bibitem [{\citenamefont {Sakurai}(1960)}]{SAKURAI19601}%
  \BibitemOpen
  \bibfield  {author} {\bibinfo {author} {\bibfnamefont {J.}~\bibnamefont
  {Sakurai}},\ }\href {\doibase https://doi.org/10.1016/0003-4916(60)90126-3}
  {\bibfield  {journal} {\bibinfo  {journal} {Annals of Physics}\ }\textbf
  {\bibinfo {volume} {11}},\ \bibinfo {pages} {1} (\bibinfo {year}
  {1960})}\BibitemShut {NoStop}%
\bibitem [{\citenamefont {Gell-Mann}\ and\ \citenamefont
  {Zachariasen}(1961)}]{Gell-Mann:1961jim}%
  \BibitemOpen
  \bibfield  {author} {\bibinfo {author} {\bibfnamefont {M.}~\bibnamefont
  {Gell-Mann}}\ and\ \bibinfo {author} {\bibfnamefont {F.}~\bibnamefont
  {Zachariasen}},\ }\href {\doibase 10.1103/PhysRev.124.953} {\bibfield
  {journal} {\bibinfo  {journal} {Phys. Rev.}\ }\textbf {\bibinfo {volume}
  {124}},\ \bibinfo {pages} {953} (\bibinfo {year} {1961})}\BibitemShut
  {NoStop}%
\bibitem [{\citenamefont {Titov}\ \emph {et~al.}(2007)\citenamefont {Titov},
  \citenamefont {Nakano}, \citenamefont {Date},\ and\ \citenamefont
  {Ohashi}}]{Titov:2007xb}%
  \BibitemOpen
  \bibfield  {author} {\bibinfo {author} {\bibfnamefont {A.~I.}\ \bibnamefont
  {Titov}}, \bibinfo {author} {\bibfnamefont {T.}~\bibnamefont {Nakano}},
  \bibinfo {author} {\bibfnamefont {S.}~\bibnamefont {Date}}, \ and\ \bibinfo
  {author} {\bibfnamefont {Y.}~\bibnamefont {Ohashi}},\ }\href {\doibase
  10.1103/PhysRevC.76.048202} {\bibfield  {journal} {\bibinfo  {journal} {Phys.
  Rev. C}\ }\textbf {\bibinfo {volume} {76}},\ \bibinfo {pages} {048202}
  (\bibinfo {year} {2007})},\ \Eprint {http://arxiv.org/abs/hep-ph/0703227}
  {arXiv:hep-ph/0703227} \BibitemShut {NoStop}%
\bibitem [{\citenamefont {Strakovsky}\ \emph {et~al.}(2015)\citenamefont
  {Strakovsky} \emph {et~al.}}]{Strakovsky:2014wja}%
  \BibitemOpen
  \bibfield  {author} {\bibinfo {author} {\bibfnamefont {I.~I.}\ \bibnamefont
  {Strakovsky}} \emph {et~al.},\ }\href {\doibase 10.1103/PhysRevC.91.045207}
  {\bibfield  {journal} {\bibinfo  {journal} {Phys. Rev. C}\ }\textbf {\bibinfo
  {volume} {91}},\ \bibinfo {pages} {045207} (\bibinfo {year} {2015})},\
  \Eprint {http://arxiv.org/abs/1407.3465} {arXiv:1407.3465 [nucl-ex]}
  \BibitemShut {NoStop}%
\bibitem [{\citenamefont {Strakovsky}\ \emph
  {et~al.}(2020{\natexlab{a}})\citenamefont {Strakovsky}, \citenamefont
  {Epifanov},\ and\ \citenamefont {Pentchev}}]{Strakovsky:2019bev}%
  \BibitemOpen
  \bibfield  {author} {\bibinfo {author} {\bibfnamefont {I.}~\bibnamefont
  {Strakovsky}}, \bibinfo {author} {\bibfnamefont {D.}~\bibnamefont
  {Epifanov}}, \ and\ \bibinfo {author} {\bibfnamefont {L.}~\bibnamefont
  {Pentchev}},\ }\href {\doibase 10.1103/PhysRevC.101.042201} {\bibfield
  {journal} {\bibinfo  {journal} {Phys. Rev. C}\ }\textbf {\bibinfo {volume}
  {101}},\ \bibinfo {pages} {042201} (\bibinfo {year} {2020}{\natexlab{a}})},\
  \Eprint {http://arxiv.org/abs/1911.12686} {arXiv:1911.12686 [hep-ph]}
  \BibitemShut {NoStop}%
\bibitem [{\citenamefont {Strakovsky}\ \emph
  {et~al.}(2020{\natexlab{b}})\citenamefont {Strakovsky}, \citenamefont
  {Pentchev},\ and\ \citenamefont {Titov}}]{Strakovsky:2020uqs}%
  \BibitemOpen
  \bibfield  {author} {\bibinfo {author} {\bibfnamefont {I.~I.}\ \bibnamefont
  {Strakovsky}}, \bibinfo {author} {\bibfnamefont {L.}~\bibnamefont
  {Pentchev}}, \ and\ \bibinfo {author} {\bibfnamefont {A.}~\bibnamefont
  {Titov}},\ }\href {\doibase 10.1103/PhysRevC.101.045201} {\bibfield
  {journal} {\bibinfo  {journal} {Phys. Rev. C}\ }\textbf {\bibinfo {volume}
  {101}},\ \bibinfo {pages} {045201} (\bibinfo {year} {2020}{\natexlab{b}})},\
  \Eprint {http://arxiv.org/abs/2001.08851} {arXiv:2001.08851 [hep-ph]}
  \BibitemShut {NoStop}%
\bibitem [{\citenamefont {Pentchev}\ and\ \citenamefont
  {Strakovsky}(2021)}]{Pentchev:2020kao}%
  \BibitemOpen
  \bibfield  {author} {\bibinfo {author} {\bibfnamefont {L.}~\bibnamefont
  {Pentchev}}\ and\ \bibinfo {author} {\bibfnamefont {I.~I.}\ \bibnamefont
  {Strakovsky}},\ }\href {\doibase 10.1140/epja/s10050-021-00364-4} {\bibfield
  {journal} {\bibinfo  {journal} {Eur. Phys. J. A}\ }\textbf {\bibinfo {volume}
  {57}},\ \bibinfo {pages} {56} (\bibinfo {year} {2021})},\ \Eprint
  {http://arxiv.org/abs/2009.04502} {arXiv:2009.04502 [hep-ph]} \BibitemShut
  {NoStop}%
\bibitem [{\citenamefont {Wang}\ \emph
  {et~al.}(2022{\natexlab{a}})\citenamefont {Wang}, \citenamefont {Zeng},\ and\
  \citenamefont {Strakovsky}}]{Wang:2022xpw}%
  \BibitemOpen
  \bibfield  {author} {\bibinfo {author} {\bibfnamefont {X.-Y.}\ \bibnamefont
  {Wang}}, \bibinfo {author} {\bibfnamefont {F.}~\bibnamefont {Zeng}}, \ and\
  \bibinfo {author} {\bibfnamefont {I.~I.}\ \bibnamefont {Strakovsky}},\ }\href
  {\doibase 10.1103/PhysRevC.106.015202} {\bibfield  {journal} {\bibinfo
  {journal} {Phys. Rev. C}\ }\textbf {\bibinfo {volume} {106}},\ \bibinfo
  {pages} {015202} (\bibinfo {year} {2022}{\natexlab{a}})},\ \Eprint
  {http://arxiv.org/abs/2205.07661} {arXiv:2205.07661 [hep-ph]} \BibitemShut
  {NoStop}%
\bibitem [{\citenamefont {Wang}\ \emph
  {et~al.}(2022{\natexlab{b}})\citenamefont {Wang}, \citenamefont {Zeng},
  \citenamefont {Wang},\ and\ \citenamefont {Zhang}}]{Wang:2022zwz}%
  \BibitemOpen
  \bibfield  {author} {\bibinfo {author} {\bibfnamefont {X.-Y.}\ \bibnamefont
  {Wang}}, \bibinfo {author} {\bibfnamefont {F.}~\bibnamefont {Zeng}}, \bibinfo
  {author} {\bibfnamefont {Q.}~\bibnamefont {Wang}}, \ and\ \bibinfo {author}
  {\bibfnamefont {L.}~\bibnamefont {Zhang}},\ }\href@noop {} {\  (\bibinfo
  {year} {2022}{\natexlab{b}})},\ \Eprint {http://arxiv.org/abs/2206.09170}
  {arXiv:2206.09170 [nucl-th]} \BibitemShut {NoStop}%
\bibitem [{\citenamefont {Han}\ \emph {et~al.}(2022)\citenamefont {Han},
  \citenamefont {Kou}, \citenamefont {Wang},\ and\ \citenamefont
  {Chen}}]{CDHan:2022cdhan}%
  \BibitemOpen
  \bibfield  {author} {\bibinfo {author} {\bibfnamefont {C.}~\bibnamefont
  {Han}}, \bibinfo {author} {\bibfnamefont {W.}~\bibnamefont {Kou}}, \bibinfo
  {author} {\bibfnamefont {R.}~\bibnamefont {Wang}}, \ and\ \bibinfo {author}
  {\bibfnamefont {X.}~\bibnamefont {Chen}},\ }\href@noop {} {\  (\bibinfo
  {year} {2022})},\ \Eprint {http://arxiv.org/abs/2210.11276} {arXiv:2210.11276
  [nucl-th]} \BibitemShut {NoStop}%
\bibitem [{\citenamefont {Strakovsky}\ \emph {et~al.}(2021)\citenamefont
  {Strakovsky}, \citenamefont {Briscoe}, \citenamefont {Pentchev},\ and\
  \citenamefont {Schmidt}}]{Strakovsky:2021vyk}%
  \BibitemOpen
  \bibfield  {author} {\bibinfo {author} {\bibfnamefont {I.~I.}\ \bibnamefont
  {Strakovsky}}, \bibinfo {author} {\bibfnamefont {W.~J.}\ \bibnamefont
  {Briscoe}}, \bibinfo {author} {\bibfnamefont {L.}~\bibnamefont {Pentchev}}, \
  and\ \bibinfo {author} {\bibfnamefont {A.}~\bibnamefont {Schmidt}},\ }\href
  {\doibase 10.1103/PhysRevD.104.074028} {\bibfield  {journal} {\bibinfo
  {journal} {Phys. Rev. D}\ }\textbf {\bibinfo {volume} {104}},\ \bibinfo
  {pages} {074028} (\bibinfo {year} {2021})},\ \Eprint
  {http://arxiv.org/abs/2108.02871} {arXiv:2108.02871 [hep-ph]} \BibitemShut
  {NoStop}%
\bibitem [{\citenamefont {Wang}\ \emph {et~al.}(2020)\citenamefont {Wang},
  \citenamefont {Evslin},\ and\ \citenamefont {Chen}}]{Wang:2019mza}%
  \BibitemOpen
  \bibfield  {author} {\bibinfo {author} {\bibfnamefont {R.}~\bibnamefont
  {Wang}}, \bibinfo {author} {\bibfnamefont {J.}~\bibnamefont {Evslin}}, \ and\
  \bibinfo {author} {\bibfnamefont {X.}~\bibnamefont {Chen}},\ }\href {\doibase
  10.1140/epjc/s10052-020-8057-9} {\bibfield  {journal} {\bibinfo  {journal}
  {Eur. Phys. J. C}\ }\textbf {\bibinfo {volume} {80}},\ \bibinfo {pages} {507}
  (\bibinfo {year} {2020})},\ \Eprint {http://arxiv.org/abs/1912.12040}
  {arXiv:1912.12040 [hep-ph]} \BibitemShut {NoStop}%
\bibitem [{\citenamefont {Kou}\ \emph {et~al.}(2022)\citenamefont {Kou},
  \citenamefont {Wang},\ and\ \citenamefont {Chen}}]{Kou:2021bez}%
  \BibitemOpen
  \bibfield  {author} {\bibinfo {author} {\bibfnamefont {W.}~\bibnamefont
  {Kou}}, \bibinfo {author} {\bibfnamefont {R.}~\bibnamefont {Wang}}, \ and\
  \bibinfo {author} {\bibfnamefont {X.}~\bibnamefont {Chen}},\ }\href {\doibase
  10.1140/epja/s10050-022-00810-x} {\bibfield  {journal} {\bibinfo  {journal}
  {Eur. Phys. J. A}\ }\textbf {\bibinfo {volume} {58}},\ \bibinfo {pages} {155}
  (\bibinfo {year} {2022})},\ \Eprint {http://arxiv.org/abs/2103.10017}
  {arXiv:2103.10017 [hep-ph]} \BibitemShut {NoStop}%
\bibitem [{\citenamefont {Ji}(1995{\natexlab{a}})}]{Ji:1995sv}%
  \BibitemOpen
  \bibfield  {author} {\bibinfo {author} {\bibfnamefont {X.-D.}\ \bibnamefont
  {Ji}},\ }\href {\doibase 10.1103/PhysRevD.52.271} {\bibfield  {journal}
  {\bibinfo  {journal} {Phys. Rev. D}\ }\textbf {\bibinfo {volume} {52}},\
  \bibinfo {pages} {271} (\bibinfo {year} {1995}{\natexlab{a}})},\ \Eprint
  {http://arxiv.org/abs/hep-ph/9502213} {arXiv:hep-ph/9502213} \BibitemShut
  {NoStop}%
\bibitem [{\citenamefont {Gryniuk}\ \emph {et~al.}(2020)\citenamefont
  {Gryniuk}, \citenamefont {Joosten}, \citenamefont {Meziani},\ and\
  \citenamefont {Vanderhaeghen}}]{Gryniuk:2020mlh}%
  \BibitemOpen
  \bibfield  {author} {\bibinfo {author} {\bibfnamefont {O.}~\bibnamefont
  {Gryniuk}}, \bibinfo {author} {\bibfnamefont {S.}~\bibnamefont {Joosten}},
  \bibinfo {author} {\bibfnamefont {Z.-E.}\ \bibnamefont {Meziani}}, \ and\
  \bibinfo {author} {\bibfnamefont {M.}~\bibnamefont {Vanderhaeghen}},\ }\href
  {\doibase 10.1103/PhysRevD.102.014016} {\bibfield  {journal} {\bibinfo
  {journal} {Phys. Rev. D}\ }\textbf {\bibinfo {volume} {102}},\ \bibinfo
  {pages} {014016} (\bibinfo {year} {2020})},\ \Eprint
  {http://arxiv.org/abs/2005.09293} {arXiv:2005.09293 [hep-ph]} \BibitemShut
  {NoStop}%
\bibitem [{\citenamefont {Ji}(1995{\natexlab{b}})}]{Ji:1994av}%
  \BibitemOpen
  \bibfield  {author} {\bibinfo {author} {\bibfnamefont {X.-D.}\ \bibnamefont
  {Ji}},\ }\href {\doibase 10.1103/PhysRevLett.74.1071} {\bibfield  {journal}
  {\bibinfo  {journal} {Phys. Rev. Lett.}\ }\textbf {\bibinfo {volume} {74}},\
  \bibinfo {pages} {1071} (\bibinfo {year} {1995}{\natexlab{b}})},\ \Eprint
  {http://arxiv.org/abs/hep-ph/9410274} {arXiv:hep-ph/9410274} \BibitemShut
  {NoStop}%
\bibitem [{\citenamefont {Buras}(1980)}]{Buras:1979yt}%
  \BibitemOpen
  \bibfield  {author} {\bibinfo {author} {\bibfnamefont {A.~J.}\ \bibnamefont
  {Buras}},\ }\href {\doibase 10.1103/RevModPhys.52.199} {\bibfield  {journal}
  {\bibinfo  {journal} {Rev. Mod. Phys.}\ }\textbf {\bibinfo {volume} {52}},\
  \bibinfo {pages} {199} (\bibinfo {year} {1980})}\BibitemShut {NoStop}%
\bibitem [{\citenamefont {Kharzeev}\ \emph {et~al.}(1999)\citenamefont
  {Kharzeev}, \citenamefont {Satz}, \citenamefont {Syamtomov},\ and\
  \citenamefont {Zinovjev}}]{Kharzeev:1998bz}%
  \BibitemOpen
  \bibfield  {author} {\bibinfo {author} {\bibfnamefont {D.}~\bibnamefont
  {Kharzeev}}, \bibinfo {author} {\bibfnamefont {H.}~\bibnamefont {Satz}},
  \bibinfo {author} {\bibfnamefont {A.}~\bibnamefont {Syamtomov}}, \ and\
  \bibinfo {author} {\bibfnamefont {G.}~\bibnamefont {Zinovjev}},\ }\href
  {\doibase 10.1007/s100529900047} {\bibfield  {journal} {\bibinfo  {journal}
  {Eur. Phys. J. C}\ }\textbf {\bibinfo {volume} {9}},\ \bibinfo {pages} {459}
  (\bibinfo {year} {1999})},\ \Eprint {http://arxiv.org/abs/hep-ph/9901375}
  {arXiv:hep-ph/9901375} \BibitemShut {NoStop}%
\bibitem [{\citenamefont {Kharzeev}(1996)}]{Kharzeev:1995ij}%
  \BibitemOpen
  \bibfield  {author} {\bibinfo {author} {\bibfnamefont {D.}~\bibnamefont
  {Kharzeev}},\ }\href {\doibase 10.3254/978-1-61499-215-8-105} {\bibfield
  {journal} {\bibinfo  {journal} {Proc. Int. Sch. Phys. Fermi}\ }\textbf
  {\bibinfo {volume} {130}},\ \bibinfo {pages} {105} (\bibinfo {year}
  {1996})},\ \Eprint {http://arxiv.org/abs/nucl-th/9601029}
  {arXiv:nucl-th/9601029} \BibitemShut {NoStop}%
\bibitem [{\citenamefont {Krutov}\ \emph {et~al.}(2016)\citenamefont {Krutov},
  \citenamefont {Polezhaev},\ and\ \citenamefont {Troitsky}}]{Krutov:2016uhy}%
  \BibitemOpen
  \bibfield  {author} {\bibinfo {author} {\bibfnamefont {A.~F.}\ \bibnamefont
  {Krutov}}, \bibinfo {author} {\bibfnamefont {R.~G.}\ \bibnamefont
  {Polezhaev}}, \ and\ \bibinfo {author} {\bibfnamefont {V.~E.}\ \bibnamefont
  {Troitsky}},\ }\href {\doibase 10.1103/PhysRevD.93.036007} {\bibfield
  {journal} {\bibinfo  {journal} {Phys. Rev. D}\ }\textbf {\bibinfo {volume}
  {93}},\ \bibinfo {pages} {036007} (\bibinfo {year} {2016})},\ \Eprint
  {http://arxiv.org/abs/1602.00907} {arXiv:1602.00907 [hep-ph]} \BibitemShut
  {NoStop}%
\bibitem [{\citenamefont {Bhagwat}\ and\ \citenamefont
  {Maris}(2008)}]{Bhagwat:2006pu}%
  \BibitemOpen
  \bibfield  {author} {\bibinfo {author} {\bibfnamefont {M.~S.}\ \bibnamefont
  {Bhagwat}}\ and\ \bibinfo {author} {\bibfnamefont {P.}~\bibnamefont
  {Maris}},\ }\href {\doibase 10.1103/PhysRevC.77.025203} {\bibfield  {journal}
  {\bibinfo  {journal} {Phys. Rev. C}\ }\textbf {\bibinfo {volume} {77}},\
  \bibinfo {pages} {025203} (\bibinfo {year} {2008})},\ \Eprint
  {http://arxiv.org/abs/nucl-th/0612069} {arXiv:nucl-th/0612069} \BibitemShut
  {NoStop}%
\bibitem [{\citenamefont {Grigoryan}\ and\ \citenamefont
  {Radyushkin}(2007)}]{Grigoryan:2007my}%
  \BibitemOpen
  \bibfield  {author} {\bibinfo {author} {\bibfnamefont {H.~R.}\ \bibnamefont
  {Grigoryan}}\ and\ \bibinfo {author} {\bibfnamefont {A.~V.}\ \bibnamefont
  {Radyushkin}},\ }\href {\doibase 10.1103/PhysRevD.76.095007} {\bibfield
  {journal} {\bibinfo  {journal} {Phys. Rev. D}\ }\textbf {\bibinfo {volume}
  {76}},\ \bibinfo {pages} {095007} (\bibinfo {year} {2007})},\ \Eprint
  {http://arxiv.org/abs/0706.1543} {arXiv:0706.1543 [hep-ph]} \BibitemShut
  {NoStop}%
\bibitem [{\citenamefont {Peskin}(1979)}]{Peskin:1979va}%
  \BibitemOpen
  \bibfield  {author} {\bibinfo {author} {\bibfnamefont {M.~E.}\ \bibnamefont
  {Peskin}},\ }\href {\doibase 10.1016/0550-3213(79)90199-8} {\bibfield
  {journal} {\bibinfo  {journal} {Nucl. Phys. B}\ }\textbf {\bibinfo {volume}
  {156}},\ \bibinfo {pages} {365} (\bibinfo {year} {1979})}\BibitemShut
  {NoStop}%
\bibitem [{\citenamefont {Kharzeev}\ \emph {et~al.}(1996)\citenamefont
  {Kharzeev}, \citenamefont {Satz}, \citenamefont {Syamtomov},\ and\
  \citenamefont {Zinovev}}]{Kharzeev:1996tw}%
  \BibitemOpen
  \bibfield  {author} {\bibinfo {author} {\bibfnamefont {D.}~\bibnamefont
  {Kharzeev}}, \bibinfo {author} {\bibfnamefont {H.}~\bibnamefont {Satz}},
  \bibinfo {author} {\bibfnamefont {A.}~\bibnamefont {Syamtomov}}, \ and\
  \bibinfo {author} {\bibfnamefont {G.}~\bibnamefont {Zinovev}},\ }\href
  {\doibase 10.1016/S0370-2693(96)01321-4} {\bibfield  {journal} {\bibinfo
  {journal} {Phys. Lett. B}\ }\textbf {\bibinfo {volume} {389}},\ \bibinfo
  {pages} {595} (\bibinfo {year} {1996})},\ \Eprint
  {http://arxiv.org/abs/hep-ph/9605448} {arXiv:hep-ph/9605448} \BibitemShut
  {NoStop}%
\bibitem [{\citenamefont {Ishikawa}\ \emph {et~al.}(2020)\citenamefont
  {Ishikawa} \emph {et~al.}}]{Ishikawa:2019rvz}%
  \BibitemOpen
  \bibfield  {author} {\bibinfo {author} {\bibfnamefont {T.}~\bibnamefont
  {Ishikawa}} \emph {et~al.},\ }\href {\doibase 10.1103/PhysRevC.101.052201}
  {\bibfield  {journal} {\bibinfo  {journal} {Phys. Rev. C}\ }\textbf {\bibinfo
  {volume} {101}},\ \bibinfo {pages} {052201} (\bibinfo {year} {2020})},\
  \Eprint {http://arxiv.org/abs/1904.02797} {arXiv:1904.02797 [nucl-ex]}
  \BibitemShut {NoStop}%
\bibitem [{\citenamefont {Klein}(1997)}]{klein1997appear}%
  \BibitemOpen
  \bibfield  {author} {\bibinfo {author} {\bibfnamefont {F.}~\bibnamefont
  {Klein}},\ }in\ \href@noop {} {\emph {\bibinfo {booktitle} {TJNAF Workshop on
  N$^{*}$ Physics,(Washington, DC)}}}\ (\bibinfo {year} {1997})\BibitemShut
  {NoStop}%
\bibitem [{\citenamefont {Seraydaryan}\ \emph {et~al.}(2014)\citenamefont
  {Seraydaryan} \emph {et~al.}}]{CLAS:2013jlg}%
  \BibitemOpen
  \bibfield  {author} {\bibinfo {author} {\bibfnamefont {H.}~\bibnamefont
  {Seraydaryan}} \emph {et~al.} (\bibinfo {collaboration} {CLAS}),\ }\href
  {\doibase 10.1103/PhysRevC.89.055206} {\bibfield  {journal} {\bibinfo
  {journal} {Phys. Rev. C}\ }\textbf {\bibinfo {volume} {89}},\ \bibinfo
  {pages} {055206} (\bibinfo {year} {2014})},\ \Eprint
  {http://arxiv.org/abs/1308.1363} {arXiv:1308.1363 [hep-ex]} \BibitemShut
  {NoStop}%
\bibitem [{\citenamefont {Dey}\ \emph {et~al.}(2014)\citenamefont {Dey},
  \citenamefont {Meyer}, \citenamefont {Bellis},\ and\ \citenamefont
  {Williams}}]{Dey:2014tfa}%
  \BibitemOpen
  \bibfield  {author} {\bibinfo {author} {\bibfnamefont {B.}~\bibnamefont
  {Dey}}, \bibinfo {author} {\bibfnamefont {C.~A.}\ \bibnamefont {Meyer}},
  \bibinfo {author} {\bibfnamefont {M.}~\bibnamefont {Bellis}}, \ and\ \bibinfo
  {author} {\bibfnamefont {M.}~\bibnamefont {Williams}} (\bibinfo
  {collaboration} {CLAS}),\ }\href {\doibase 10.1103/PhysRevC.89.055208}
  {\bibfield  {journal} {\bibinfo  {journal} {Phys. Rev. C}\ }\textbf {\bibinfo
  {volume} {89}},\ \bibinfo {pages} {055208} (\bibinfo {year} {2014})},\
  \bibinfo {note} {[Addendum: Phys.Rev.C 90, 019901 (2014)]},\ \Eprint
  {http://arxiv.org/abs/1403.2110} {arXiv:1403.2110 [nucl-ex]} \BibitemShut
  {NoStop}%
\bibitem [{\citenamefont {Yang}\ \emph {et~al.}(2018)\citenamefont {Yang},
  \citenamefont {Liang}, \citenamefont {Bi}, \citenamefont {Chen},
  \citenamefont {Draper}, \citenamefont {Liu},\ and\ \citenamefont
  {Liu}}]{Yang:2018nqn}%
  \BibitemOpen
  \bibfield  {author} {\bibinfo {author} {\bibfnamefont {Y.-B.}\ \bibnamefont
  {Yang}}, \bibinfo {author} {\bibfnamefont {J.}~\bibnamefont {Liang}},
  \bibinfo {author} {\bibfnamefont {Y.-J.}\ \bibnamefont {Bi}}, \bibinfo
  {author} {\bibfnamefont {Y.}~\bibnamefont {Chen}}, \bibinfo {author}
  {\bibfnamefont {T.}~\bibnamefont {Draper}}, \bibinfo {author} {\bibfnamefont
  {K.-F.}\ \bibnamefont {Liu}}, \ and\ \bibinfo {author} {\bibfnamefont
  {Z.}~\bibnamefont {Liu}},\ }\href {\doibase 10.1103/PhysRevLett.121.212001}
  {\bibfield  {journal} {\bibinfo  {journal} {Phys. Rev. Lett.}\ }\textbf
  {\bibinfo {volume} {121}},\ \bibinfo {pages} {212001} (\bibinfo {year}
  {2018})},\ \Eprint {http://arxiv.org/abs/1808.08677} {arXiv:1808.08677
  [hep-lat]} \BibitemShut {NoStop}%
\bibitem [{\citenamefont {Accardi}\ \emph {et~al.}(2016)\citenamefont {Accardi}
  \emph {et~al.}}]{Accardi:2012qut}%
  \BibitemOpen
  \bibfield  {author} {\bibinfo {author} {\bibfnamefont {A.}~\bibnamefont
  {Accardi}} \emph {et~al.},\ }\href {\doibase 10.1140/epja/i2016-16268-9}
  {\bibfield  {journal} {\bibinfo  {journal} {Eur. Phys. J. A}\ }\textbf
  {\bibinfo {volume} {52}},\ \bibinfo {pages} {268} (\bibinfo {year} {2016})},\
  \Eprint {http://arxiv.org/abs/1212.1701} {arXiv:1212.1701 [nucl-ex]}
  \BibitemShut {NoStop}%
\bibitem [{\citenamefont {Chen}(2018)}]{Chen:2018wyz}%
  \BibitemOpen
  \bibfield  {author} {\bibinfo {author} {\bibfnamefont {X.}~\bibnamefont
  {Chen}},\ }\bibfield  {booktitle} {\emph {\bibinfo {booktitle} {{Proceedings,
  26th International Workshop on Deep Inelastic Scattering and Related Subjects
  (DIS 2018): Port Island, Kobe, Japan, April 16-20, 2018}}},\ }\href {\doibase
  10.22323/1.316.0170} {\bibfield  {journal} {\bibinfo  {journal} {PoS}\
  }\textbf {\bibinfo {volume} {DIS2018}},\ \bibinfo {pages} {170} (\bibinfo
  {year} {2018})},\ \Eprint {http://arxiv.org/abs/1809.00448} {arXiv:1809.00448
  [nucl-ex]} \BibitemShut {NoStop}%
\bibitem [{\citenamefont {Chen}\ \emph {et~al.}(2020)\citenamefont {Chen},
  \citenamefont {Guo}, \citenamefont {Roberts},\ and\ \citenamefont
  {Wang}}]{Chen:2020ijn}%
  \BibitemOpen
  \bibfield  {author} {\bibinfo {author} {\bibfnamefont {X.}~\bibnamefont
  {Chen}}, \bibinfo {author} {\bibfnamefont {F.-K.}\ \bibnamefont {Guo}},
  \bibinfo {author} {\bibfnamefont {C.~D.}\ \bibnamefont {Roberts}}, \ and\
  \bibinfo {author} {\bibfnamefont {R.}~\bibnamefont {Wang}},\ }\href {\doibase
  10.1007/s00601-020-01574-0} {\bibfield  {journal} {\bibinfo  {journal} {Few
  Body Syst.}\ }\textbf {\bibinfo {volume} {61}},\ \bibinfo {pages} {43}
  (\bibinfo {year} {2020})},\ \Eprint {http://arxiv.org/abs/2008.00102}
  {arXiv:2008.00102 [hep-ph]} \BibitemShut {NoStop}%
\bibitem [{\citenamefont {Anderle}\ \emph {et~al.}(2021)\citenamefont {Anderle}
  \emph {et~al.}}]{Anderle:2021wcy}%
  \BibitemOpen
  \bibfield  {author} {\bibinfo {author} {\bibfnamefont {D.~P.}\ \bibnamefont
  {Anderle}} \emph {et~al.},\ }\href@noop {} {\  (\bibinfo {year} {2021})},\
  \Eprint {http://arxiv.org/abs/2102.09222} {arXiv:2102.09222 [nucl-ex]}
  \BibitemShut {NoStop}%
\end{thebibliography}%

\end{document}